\documentclass[nofootinbib,prl,article,epsf,psfig,12pt]{revtex4}
\usepackage{graphicx}

\begin{document}

\title{Low temperature thermodynamic properties of a polarized Fermi gas in a quartic trap}

\author{Mark DelloStritto and Theja N. De Silva}
\affiliation{Department of Physics, Applied Physics and Astronomy,
The State University of New York at Binghamton, Binghamton, New York
13902, USA.}

\date{\today}

\begin{abstract}
We study low temperature thermodynamic properties of a polarized Fermi gas trapped in a quartic anharmonic potential.
We use a semi classical approximation and a low temperature series expansion method to derive analytical expressions for various thermodynamic quantities.
These quantities include the total energy, particle number, and
heat capacity for both positive and negative anharmonic confinements.

\textbf{Keywords:} anharmonic potential, chemical potential, Fermi-Dirac statistics, Fermi energy, fermions, polarized fermi gas, statistical mechanics, thermodynamics, Thomas-Fermi semi-classical approximation, specific heat
\end{abstract}

\maketitle

\section{I. Introduction}

Since its realization in dilute atomic gases, superfluidity of
alkali atoms has been studied extensively in various externally controllable environments~\cite{ex}.
As a controllable parameter, the external trapping potential of a cold atomic system plays an important constrained
role in current experimental setups. Most of the experiments are carried out in either magnetic or optical harmonic traps.
A negative, but small quartic term is always present with the Gaussian optical potentials in current
experimental setups. For the case of rapidly rotating
Fermi gasses in harmonic traps, the force due to the trapping frequency almost
balances the centrifugal force and the atomic cloud
spreads in the plane perpendicular to the rotation axis. Therefore, an added positive quartic potential term
ensures the stability of the fast rotating regime~\cite{qu, dali}. Adding a positive
quartic term is very important for a two-component superfluid Fermi system
where the system is expected to be in the quantum
Hall regime at the limit of extreme larger rotations~\cite{fqh}.

Although the interaction between the atoms is extremely important
in a system, the problems are made tractable and the essential
physics is retained by assuming fully
polarized non-interacting atoms. Experimentally such a
system can be prepared by trapping a single component
(single hyperfine spin state) of atomic species. Because of the
large spin relaxation time of the atoms compared to
the other experimental time scales, single species can be retained over the entire time of the
experiments. For a system of fully polarized Fermi
gas, the \emph{s}-wave scattering is prohibited due to the
Pauli exclusion principle, while higher order partial wave
scattering is negligible at low temperatures. Therefore, we model the system as
a collection of non-interacting atoms obeying Fermi-Dirac statistics confined
by quadratic and quartic potentials given as

\begin{eqnarray}\label{trap}
V(r,z) &=& \frac{1}{2}M\omega_r^2r^2 + \frac{1}{2}M\omega_z^2z^2
+ \gamma r^4 \nonumber \\
 &=& br^2 + b_z z^2 + \gamma r^4
\end{eqnarray}

\noindent where $M$ is the atom mass, $\omega_i$'s are the harmonic
trapping frequencies ($i = r, z)$, and $r^2 = x^2+y^2$. Notice that
we are using the cartesian coordinate system with coordinates $(x,
y, z)$. We use Thomas-Fermi semiclassical approximation to derive
various low temperature thermodynamic properties as a function of
anharmonic confinement $\gamma$ for \emph{both} positive and
negative values. Fast growing progress in the current experiments to
manipulate quantum many body states in rotating Fermi gases
motivated us to study the effect of positive $\gamma$
values~\cite{dali}. Experiments in harmonic traps are limited by the
loss of confinement. This loss is due to centrifugal force when the
rotation frequency approaches the axial harmonic trapping frequency.
One can overcome this loss by adding a quartic term to the trap
potential as already implemented in
experiments~\cite{dali}\footnote{We do not consider rotation in the
present work. Fast rotating Fermi gas can be linked to one of the
very interesting phenomena; integer quantum Hall effect.}. On the
other hand, the small negative quartic term present with the
Gaussian optical potential can be canceled by adding an extra
quartic term. We will show later, that the minimum negative $\gamma$
value at which the atomic cloud is confined depends on the total
number of atoms present in the system. Thermodynamic properties of
Fermi gases confined in a harmonic trap have been studied in
Ref.~\cite{fermi}. In a recent work, interplay between rotation and
anharmonicity of a Fermi gas has been studied extensively in
Ref.~\cite{howe}. In the superfluid phase of a two-component gas,
the effect of anharmonicities on breathing mode frequencies in the
Bardeen-Cooper-Schriefer- BEC crossover regime has been presented in
Ref.~\cite{theja}. Thermodynamic properties of a Bose gas confined
in a quartic potential trap can be found in ref.~\cite{bose}.

This paper is organized as follows. In section II, we present the
derivation of thermodynamic quantities within our framework of semiclassical approximation.
In section III, we present our results discussing how these thermodynamic quantities depend on trap potential parameters.
Finally in section IV, we draw our conclusions.

\section{II. Formalism}

We consider a polarized Fermi atomic system trapped in an external potential which has both harmonic and radial quartic
components given in Eq. \ref{trap}. For the atoms at thermal equilibrium,
the average number of atoms in the single particle state $|\alpha >$ with energy $\epsilon_\alpha$ is given

\begin{eqnarray}\label{avn}
n_\alpha = \frac{1}{e^{\beta (\epsilon_\alpha-\mu)}+1}
\end{eqnarray}

\noindent where $\mu$ is the chemical potential and $\beta = 1/(k_BT)$, with $k_B$ the Boltzmann
constant and $T$ the temperature. The average number $N = \sum_\alpha g n_\alpha$ of atoms of the system
fixes the chemical potential. The total energy $U = \sum_\alpha g \epsilon_\alpha n_\alpha$, where the
spin degenerate factor $g = 1$ because the fermions are polarized. When the number of atoms in the system is large and the average potential energy of atoms in the trap is much smaller than the kinetic energy of the atoms, Thomas-Fermi semiclassical
approximation can be used. The semiclassical approximation means that, instead of $\epsilon_\alpha$,
we can use the classical single-particle phase space energy $\epsilon (r, p) = p^2/(2M) + V(r,z)$. In general,
at high enough temperatures, the particles are classical so that the semiclassical approximation is valid. In the present paper, we
are considering a spin polarized Fermi system. As a result of the Pauli exclusion principle, when the temperature goes to zero,
Fermi atoms fill every energy level up to the Fermi energy $E_F$ (the energy of the highest occupied state). If the system has large number of atoms,
then the Fermi energy $E_F$ of the system is larger than the single atom ground state
energy\footnote{For a system with large enough atom numbers, majority of atoms will have average energy in the order of $E_F$.}. Therefore, a spin-polarized Fermi gas
can be well described in the semiclassical approximation even at very low temperatures. This is drastically different
for bosons where thermal energy $k_BT$ has to be compared with the ground state energy.

Using the quantum elementary volume of the single-particle phase space, $h^3$, where $h$ is the Plank constant,
the total number of atoms and the energy of the system have the forms

\begin{eqnarray}\label{number1}
N = \frac{g}{h^3}\int\frac{d^3rd^3p}{Z^{-1}e^{\beta \epsilon(r, p)} + 1}
\end{eqnarray}

\noindent and
\begin{eqnarray}\label{energy1}
U = \frac{g}{h^3}\int\frac{\epsilon(r,p)d^3rd^3p}{Z^{-1}e^{\beta \epsilon(r, p)} + 1}.
\end{eqnarray}

\noindent Here $Z = e^{\beta \mu}$. Writing the volume of the momentum element
$d^3p = 4\pi p^2 dp$ and changing of variables $\beta p^2/(2M)\rightarrow x$, these two equations can be written in the forms,

\begin{eqnarray}\label{number2}
N = \frac{g}{h^3}\biggr(\frac{2M\pi}{\beta}\biggr)^{3/2}\int d^3r f_{3/2}(z_r)
\end{eqnarray}

\noindent and

\begin{eqnarray}\label{energy2}
U = \frac{3g}{2h^3}(2M\pi)^{3/2} \frac{1}{\beta^{5/2}}\int d^3r f_{5/2}(z_r) + \frac{g}{h^3}\biggr(\frac{2M\pi}{\beta}\biggr)^{3/2}\int d^3r V(r,z)f_{3/2}(z_r).
\end{eqnarray}

\noindent Here  $z^{-1}_r = Z^{-1}e^{\beta V(r,z)}$,  and the Fermi integral $f_l(z_r)$ is given by

\begin{eqnarray}\label{fint}
f_l (z_r) = \frac{1}{\Gamma(l)}\int^\infty_0\frac{x^{l-1}dx}{z_r^{-1}e^x+1}
\end{eqnarray}

\noindent where $\Gamma(l) = \int^\infty_0 t^{l-1}e^{-t} dt$ is the Gamma function. For low temperatures (or large $z_r$ values), we expand $f_l (z_r)$ as an asymptotic series-commonly known as Sommerfeld's lemma~\cite{pathria},

\begin{eqnarray}\label{gamma}
f_l (z_r) = \frac{\ln z_r}{\Gamma(l+1)}\biggr [1+l(l-1)\frac{\pi^2}{6}\frac{1}{(\ln z_r)^2}+l(l-1)(l-2)(l-3)\frac{7\pi^2}{360}\frac{1}{(\ln z_r)^4}+...\biggr ].
\end{eqnarray}

\noindent Writing the volume element $\int d^3r =
2\int_0^{z_0}dz\int_0^{r_\perp(z)}2\pi r_\perp dr_\perp$, with the
edges\footnote{Effective chemical potential $\mu-V(r,z)$ monotonically decreases from center to the edge of the trap. So that the edge of the trap is determined by the condition $\mu-V(r,z) = 0$.} of the atomic cloud $r_\perp^2(z) =
\sqrt{b^2/(4\gamma^2)+(\mu-b_zz^2)/\gamma}-b/(2\gamma)$ and $z_0 =
\sqrt{\mu/b_z}$ [note $b$ and $b_z$ are defined in Eq~(\ref{trap})],
we can perform the spatial integration to obtain analytical
expressions for the number of atoms and the total energy;

\begin{eqnarray}\label{number3}
N = \frac{g\lambda}{\sqrt{\pi |\tilde{\gamma}}|}\biggr[\frac{4}{3\pi}\tilde{\mu}^{5/2}I_{N1}(A) + \frac{\pi^{3/2}}{6}\tilde{\mu}^{1/2}\tilde{T}^2I_{N2}(A) + {\cal O}(\tilde{T}^3) \biggr]
\end{eqnarray}

\noindent and
\begin{eqnarray}\label{energy3}
\tilde{E} = \frac{g\lambda}{\sqrt{|\tilde{\gamma}|}} \tilde{\mu}^{7/2}I_{E1}(A) + \frac{g\pi\lambda}{\sqrt{|\tilde{\gamma}|}} \tilde{\mu}^{3/2}\tilde{T}^2I_{E2}(A) +  {\cal O}(\tilde{T}^3)
\end{eqnarray}

\noindent where we have defined the dimensionless variables
$\tilde{E} \equiv U/(\hbar \omega_r)$, aspect ratio\footnote{The
aspect ratio of a thermal cloud is the ratio of the radial to axial
size in a harmonic potential.} of the harmonic trap $\lambda \equiv
\omega_r/\omega_z$, $\tilde{\gamma} \equiv \hbar
\gamma/(M^2\omega_r^3)$, $\tilde{\mu}\equiv \mu/(\hbar \omega_r)$,
$\tilde{T} \equiv k_BT/(\hbar \omega_r)$, and $A \equiv (1/16)
[1/(|\tilde{\gamma}| \tilde{\mu})]$. The dimensionless functions
$I_{N1}(x)$, $I_{N2}(x)$, $I_{E1}(x) = 4I_{U1}(x)/5 +
4[I_{U3}(x)+I_{U4}(x)]/3$, and $I_{E2}(x) = I_{U2}(x)/2 + [I_{U5}(x)
+ I_{U6}(x)]/6$ are defined below. The upper sign is for positive values of $\gamma$ while the lower sign is for negative values of $\gamma$.

\begin{eqnarray}\label{IN1}
I_{N1}(x) = \frac{1}{80} \biggr(\pm 8 (\pm 1 + x)^{5/2} \pi \mp \sqrt{x} (15 \pm 20 x + 8 x^2) \pi\biggr)
\end{eqnarray}

\begin{eqnarray}\label{IN2}
I_{N2}(x) = \pm \frac{\sqrt{x} \pi}{2} \pm \frac{\sqrt{\pm 1 +x} \pi}{2}
\end{eqnarray}

\begin{eqnarray}\label{IU1}
I_{U1}(x) = \frac{1}{96} \biggr(\mp \frac{15}{16} \sqrt{x} \pi  \left(8 x^2\pm 16 x+11\right)\mp \frac{3}{112} \sqrt{x} \left(\pm 256 x^3+616 x^2\pm 560 x+175 \right) \pi \nonumber \\
+\frac{48 \left(x^{9/2}\pm 4 x^{7/2}+6 x^{5/2}\pm 4 x^{3/2}+\sqrt{x}\biggr) \pi  \right)}{7
   \sqrt{x} \sqrt{x\pm 1}}\biggr)
\end{eqnarray}

\begin{eqnarray}\label{IU2}
I_{U2}(x) = \frac{1}{4} \left(\frac{2 \left(x^{5/2}\pm 2 x^{3/2}+\sqrt{x}\right) \pi }{3 \sqrt{x} \sqrt{x\pm 1}}-\frac{1}{3} \sqrt{x} (2 x\pm 3) \pi \right)
\end{eqnarray}

\begin{eqnarray}\label{IU3}
I_{U3}(x) = \frac{1}{280} \biggr(24 x^{7/2}-24 \sqrt{x\pm 1} x^3\pm 56 x^{5/2}\mp 44 \sqrt{x\pm 1} x^2+35 x^{3/2} \nonumber \\
-16 \sqrt{x\pm 1} x\pm 4 \sqrt{x\pm 1}\biggr) \pi
\end{eqnarray}

\begin{eqnarray}\label{IU4}
I_{U4}(x) = \frac{1}{560} \left(8 (x\pm 1)^{7/2} \pi -\frac{1}{2} \sqrt{x} \left(16 x^3\pm 56 x^2+70 x\pm 35\right) \pi \right)
\end{eqnarray}

\begin{eqnarray}\label{IU5}
I_{U5}(x) = \frac{\left(-2 x^2+2 \sqrt{x\pm 1} x^{3/2}\mp x+1\right) \pi }{6 \sqrt{x\pm 1}}
\end{eqnarray}

\begin{eqnarray}\label{IU6}
I_{U6}(x) =\frac{2\mp 3\sqrt{x(x\pm 1)}\pm 4x-2\sqrt{x\pm 1}x^{3/2}+2x^2}{12 \sqrt{x\pm 1}}.
\end{eqnarray}

\noindent The recurring $\pm$ sign originates from the sign of $\gamma$ in the integrals. Finally,  we derive heat capacity $C = (\partial U/\partial T)|_N$ of the system using Eq.~(\ref{energy3}).
First, we write

\begin{eqnarray}
\frac{\partial U}{\partial T}\biggr|_N &=& \frac{k_B}{\hbar \omega_r}\frac{\partial U}{\partial \tilde{T}}\biggr |_N \nonumber \\
&=& \frac{dU}{d\tilde{T}}+\frac{\partial U}{\partial \tilde{\mu}}\frac{\partial \tilde{\mu}}{\partial \tilde{T}}.
\end{eqnarray}

\noindent As the number of atoms in the trap is fixed, using the condition $\partial N/\partial \tilde{\mu} =0$, we find $\partial \tilde{\mu}/\partial \tilde{T}$ to obtain the heat capacity,

\begin{eqnarray}\label{sheat}
\frac{C}{k_B} = \frac{2g\pi \lambda \tilde{\mu}^{3/2}}{\sqrt{|\tilde{\gamma}|}}\tilde{T}I_{E2}(A)
-\frac{\pi^2}{10}\frac{\tilde{T}}{\tilde{\mu}}\biggr(\frac{7g\lambda \tilde{\mu}^{5/2}}{2\pi \sqrt{|\tilde{\gamma}|}}I_{E1}(A)-\frac{g\lambda \tilde{\mu}^{3/2}}{16\pi |\tilde{\gamma}|^{3/2}}\frac{\partial I_{E1}(A)}{\partial A}\biggr)\frac{I_{N2}(A)}{I_{N1}(A)} + {\cal O}(\tilde{T}^3)
\end{eqnarray}

\section{III. results}

\begin{figure}
\includegraphics[width=\columnwidth]{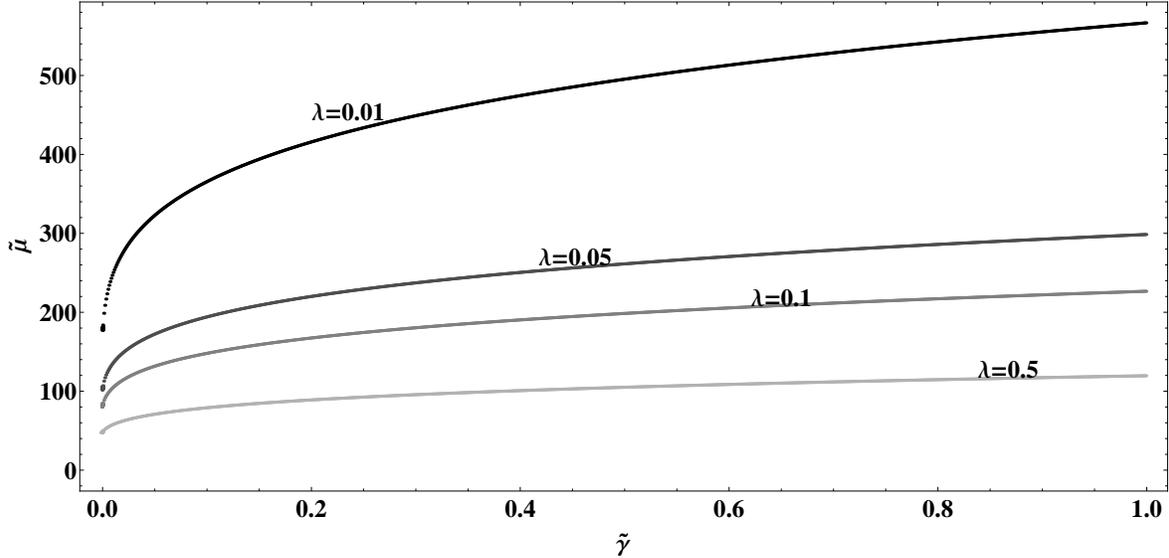}
\caption{Chemical potential as a function of anharmonic term
$\gamma$ for fixed number of atmos $N = 1.0 \times 10^4$. We fixed
the temperature to be $k_BT = \hbar\omega_r$. Different lines
represent different aspect ratios $\lambda$.} \label{mu}
\end{figure}

\begin{figure}
\includegraphics[width=\columnwidth]{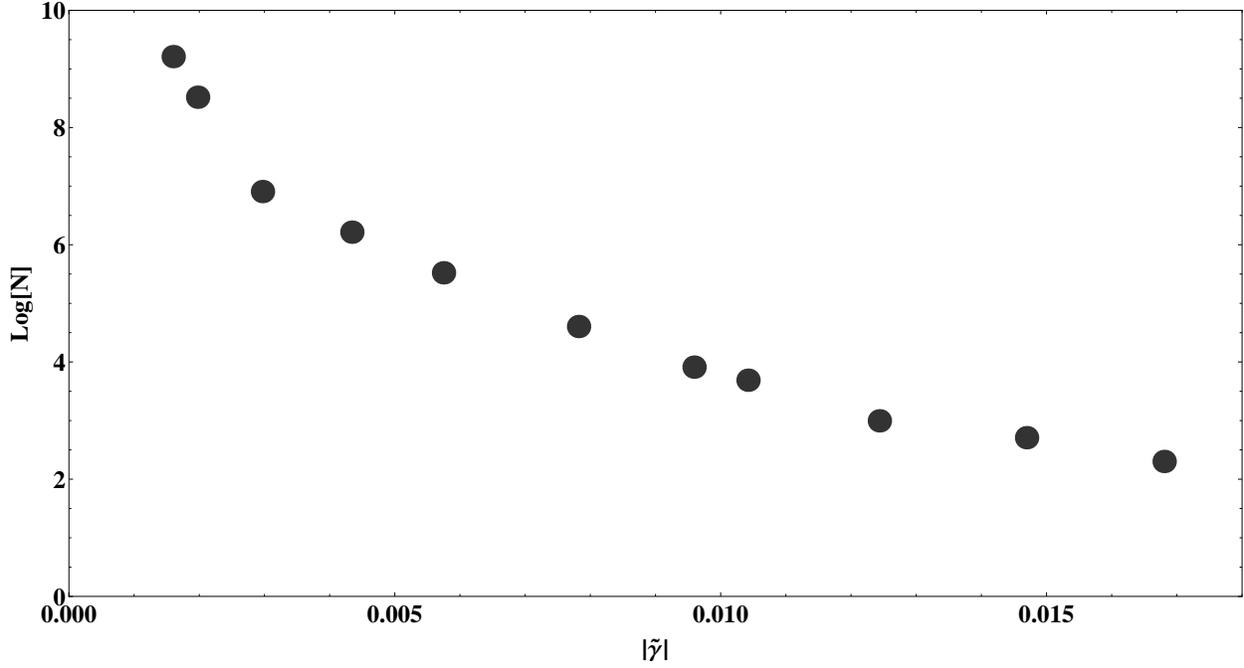}
\caption{The critical negative anharmonic dependence on number of atoms in the trap for $k_BT = \hbar\omega_r$ and $\lambda$ =1.
The atomic cloud is unstable below this critical value.} \label{unk}
\end{figure}

\begin{figure}
\includegraphics[width=\columnwidth]{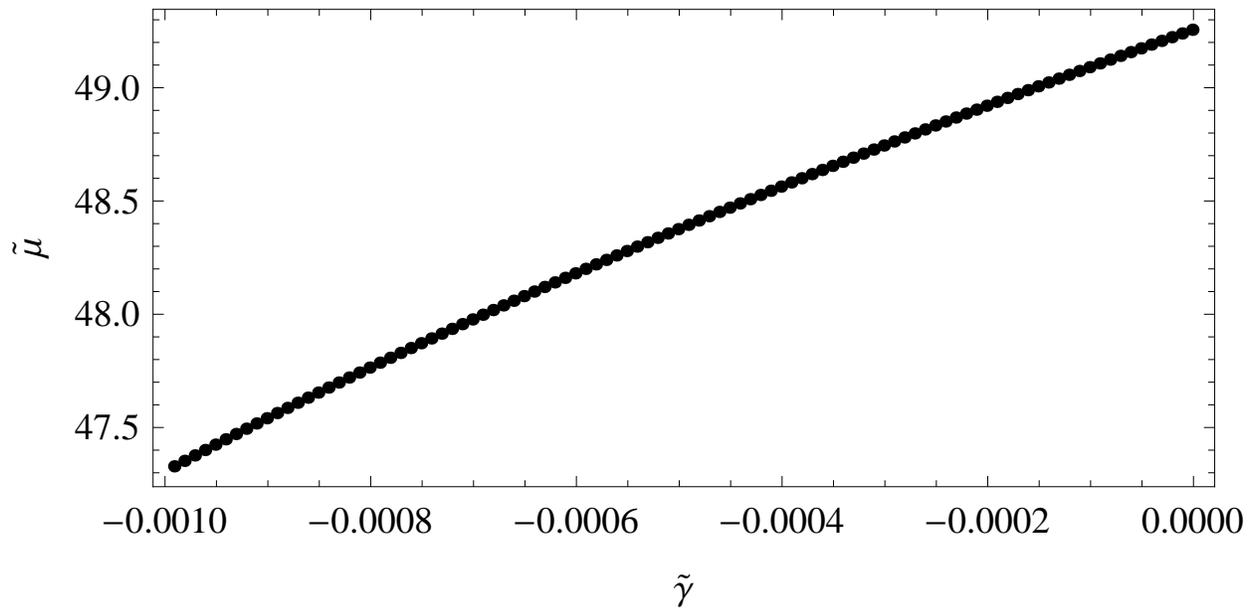}
\caption{Chemical potential for negative anharmonic potentials. We fixed the temperature,
number of atoms, and aspect ratio to be $k_BT = \hbar\omega_r$, $N = 1.0 \times 10^4$, and $\lambda = 0.5$.
For $\tilde{\gamma} < -0.001$, the system is unstable.} \label{muN}
\end{figure}

In this section we present thermodynamical properties as a function
of temperature and trap potential parameters. For our calculations,
we use a large number of atoms ($1.0 \times 10^4$) to ensure the
validity of our semi-classical treatment. First we solve
Eq.~\ref{number2} for $\tilde{\mu}$ for given $N$, $T$, and
$\lambda$. As can be seen from FIG.~\ref{mu}, while the chemical
potential increases with the anharmonic term, it decreases as the
aspect ratio($\lambda$) increases. As more energy is needed to
excite the fermions in low aspect ratio traps, the chemical
potential in low aspect ratio traps is larger than that of higher
aspect ratio traps. As we are using large number of atoms for our
calculations, anharmonic term cannot be too negative. If the
anharmonic term is too negative relative to the harmonic
confinement, the trapped atoms cannot be confined as the potential
is not bounded from below. The value of the anharmonicity at the
point where it dominates over the harmonic confinement will be
called critical anharmonicity hereafter. This critical anharmonicity
depends on the number of atoms in the trap. For a set of
representative parameters, this critical anharmonic value as a
function of atom number is given in FIG.~\ref{unk}. Chemical
potential for a range of negative anharmonic potential is given in
FIG.~\ref{muN}.

\begin{figure}
\includegraphics[width=\columnwidth]{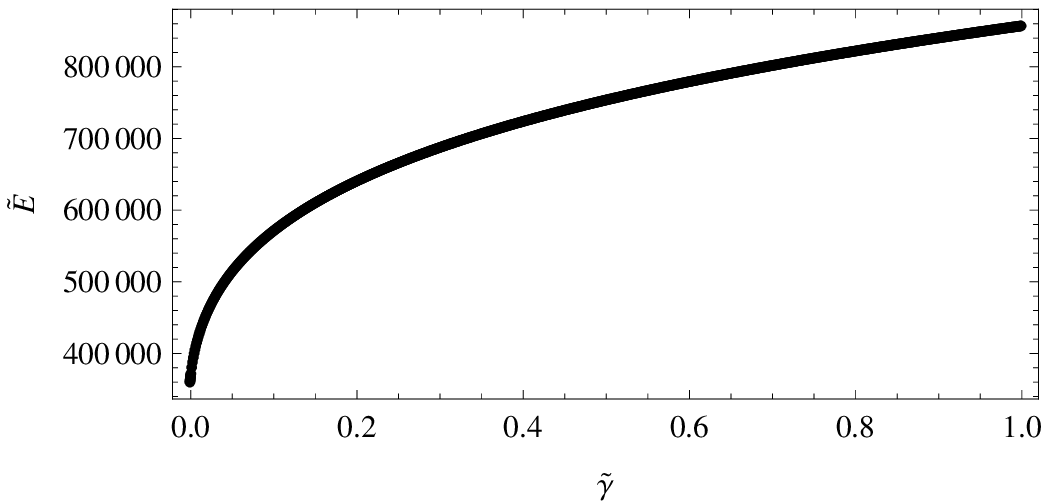}
\caption{Total energy as function of anharmonic term $\gamma$ for
fixed number of atmos $N = 1.0 \times 10^4$. We fixed the
temperature to be $k_BT = \hbar\omega_r$ and the aspect ratio
$\lambda$ = 0.5. For these selected parameters, absolute value of
critical $|\gamma|$ is 0.001.} \label{e}
\end{figure}

In FIG.~\ref{e}, we plot the total energy as a function of anharmonic term. For given values of atomic
number $N$, temperature $T$, and aspect ration $\lambda$, we first calculate the chemical potential
from Eq.~\ref{number3}, and then we use this chemical potential to calculate the energy from Eq.~\ref{energy3}. As can be seen in FIG.~\ref{e},
for  fixed number of atoms and fixed aspect ratio $\lambda$, dimensionless energy increases as we increase
the anharmonic term $\gamma$. This is because of the trapping potential energy increases with increasing $\gamma$.

\begin{figure}
\includegraphics[width=\columnwidth]{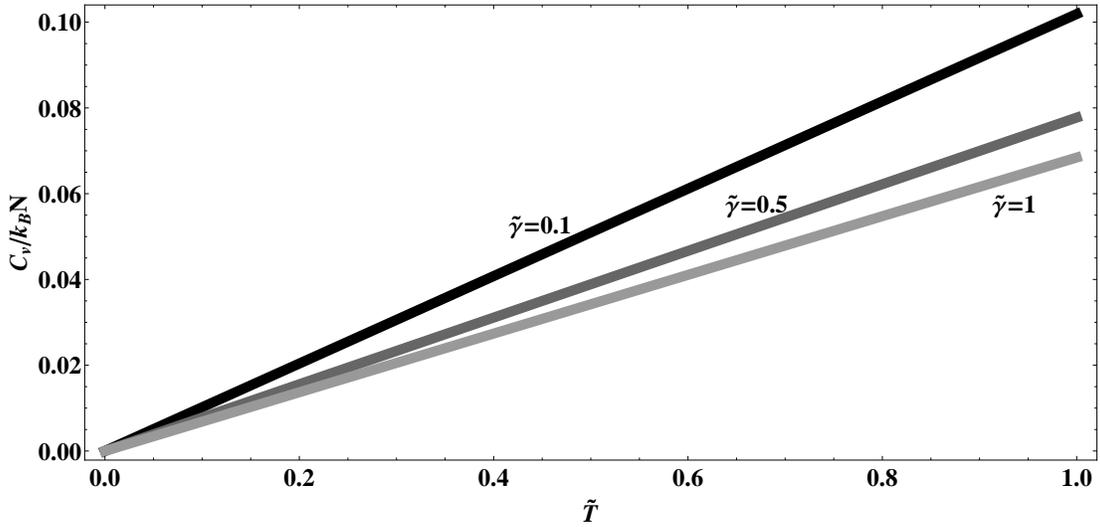}
\caption{Specific heat as function of temperature for
fixed number of atmos $N = 1.0 \times 10^4$. We fixed the aspect ratio to be $\lambda = 0.5$.} \label{sph1}
\end{figure}

\begin{figure}
\includegraphics[width=\columnwidth]{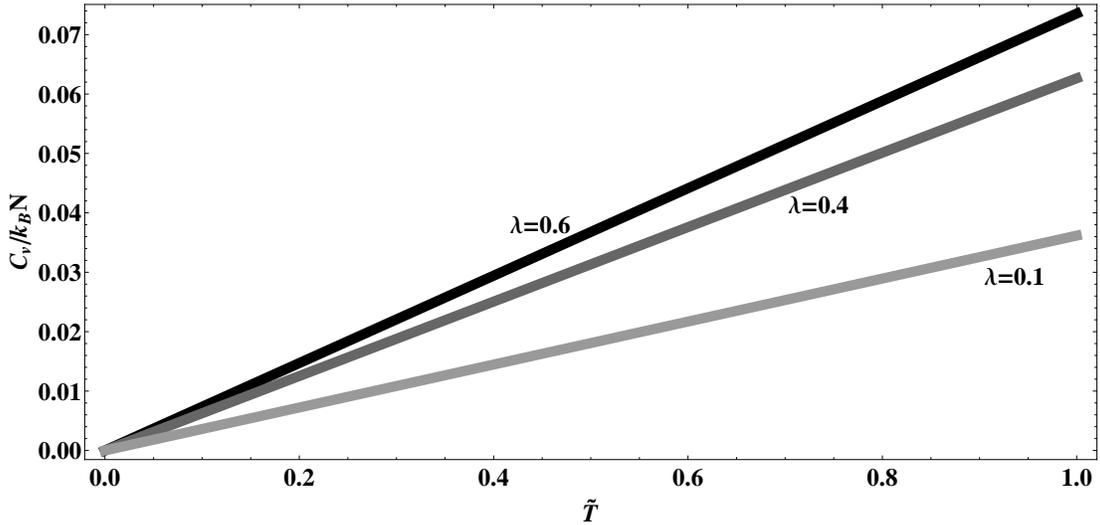}
\caption{Specific heat as function of temperature for
fixed number of atmos $N = 1.0 \times 10^4$ and $\tilde{\gamma}$ = 0.5.} \label{sph2}
\end{figure}

\begin{figure}
\includegraphics[width=\columnwidth]{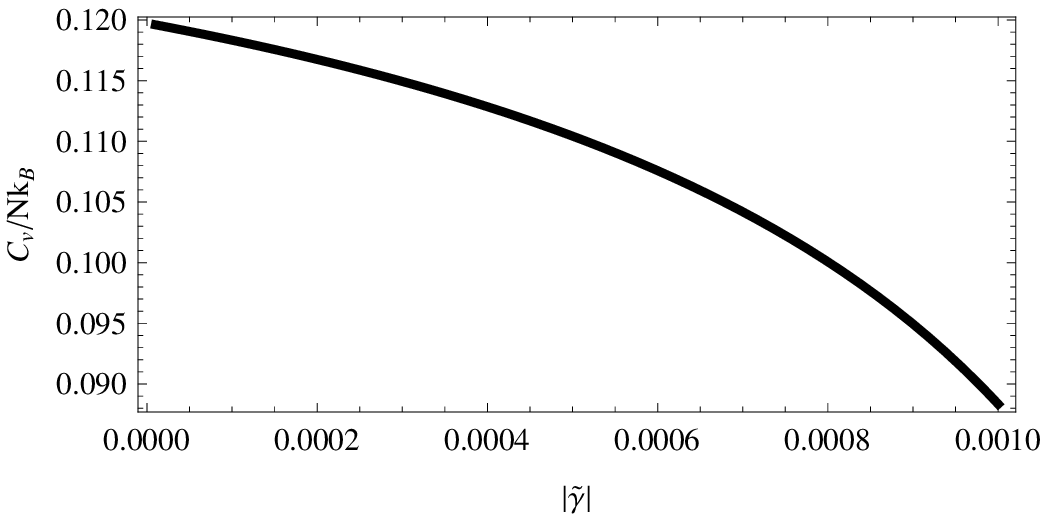}
\caption{Specific heat as function of $\gamma$ for negative $\gamma$'s. We fixed the number of atmos $N = 1 \times 10^4$. We fixed the
$k_BT = \hbar\omega_r$.} \label{sph3}
\end{figure}

The low temperature behavior of the heat capacity calculated from Eq. (\ref{sheat}) is shown in
FIGS.~\ref{sph1}-\ref{sph3}. The temperature dependence of the heat capacity for different values of $\gamma$ and $\lambda$
is shown in FIG.~\ref{sph1} and FIG.~\ref{sph2}. As we have considered up to the linear order in temperature, the specific heat is
linear with temperature, however, the slope change significantly as we change the values of $\gamma$ or $\lambda$. The heat capacity of
the polarized Fermi gas increases monotonically as the temperature increases. This behavior is completely different from a
Bose gas~\cite{cb} where the heat capacity shows a maximum at supefluid-normal phase transition point. As can be seen from the figures,
the heat capacity decreases with increasing anharmonic term $\gamma$. This decrease is due to the confinement effects which can be clearly
seen from Eq.~\ref{sheat}.

\section{IV. Conclusions}

We have considered a spin-polarized Fermi gas trapped in an anharmonic potential. Applying a semiclassical treatment to
the particle number and energy, and using a low-temperature series expansion method, we
have derived analytical expressions for several thermodynamic quantities. We have discussed the trap parameter
dependence on these quantities and the results were presented as a function of anharmonic trap parameter
and temperature. The results of our paper can be tested in a lab by a similar experimental procedure carried out by
Bretin \emph{et al}.~\cite{dali} for Bose-Einstein condensates in a quartic trap.

In the present work, we considered fully polarized fermions. At low temperatures and low densities,
\emph{s}-wave short-range interactions are prohibited and higher wave scattering are strongly suppressed. The most natural extension of
this work would be the inclusion of interaction. This can be done by adding another Zeeman state of atoms and consider the
system as a two component fermions. A particular interesting regime is so called uintarity regime, where the scattering length
between atoms is infinite. This regime is located near the Feshbach resonance~\cite{fb}. At this point, the interaction parameter
disappears from the physical quantities. Therefore, all thermodynamic quantities are universal function of the dimensionless
temperature $k_BT/\epsilon_F$, where $\epsilon_F$ is the Fermi energy~\cite{ho}. However, as there is no small parameter,
it is difficult to reliably calculate thermodynamic quantities close the the unitarity regime. Therefore, one has to rely on good
experimental data of energy, heat capacity, etc.~\cite{duke}. Then by comparing experimental data at unitarity regime with
non-interacting calculation of thermodynamic quantities allows one to find the scaling factors. An accurate
determination of these scaling factors of unitary Fermi gases is one of the challenging and
fascinating theoretical problem at the moment~\cite{mc}.

\end{document}